\documentclass[preprint]{aastex7}
\usepackage{xspace}
\usepackage{amssymb,amsfonts,amsmath}
\usepackage{comment}
\usepackage[hyphens]{url}

\newcommand{\dMdp}{\frac{d \mathrm{M}}{d\rho}} 
\newcommand{\dMinline}{$d \mathrm{M} / d \rho$\xspace}
\newcommand{\av}{$A_\mathrm{V}$\xspace}
\newcommand{\m}{\ {\rm m}}
\newcommand{\s}{\ {\rm s}}

\newcommand{\K}{\ {\rm K}}

\newcommand{\ee}[1]{\times 10^{#1}}

\newcommand{\hinv}{$h^{-1}\,$}
\newcommand{\x}{$\times$\xspace}
\newcommand{\mgII}{\ion{Mg}{2}\xspace}

\newcommand\jpl{{Jet Propulsion Laboratory, California Institute of Technology, 4800 Oak Grove Drive, Pasadena, CA, USA}}
\newcommand\northeastern{{Department of Physics, Northeastern University, 360 Huntington Ave, Boston, MA USA}}
\newcommand\cnrs{{Universit\'{e} Paris Cit\'{e}, CNRS, Astroparticule et Cosmologie, 75013 Paris, France}}

\begin{document}

\title{A Detection of Circumgalactic Dust at Megaparsec Scales with Maximum Likelihood Estimation}

\author[0000-0002-9883-7460]{Jacqueline E.\ McCleary}
\affiliation{\northeastern}
\email[show]{j.mccleary@northeastern.edu}  

\author[0000-0002-9378-3424]{Eric M. Huff}
\affiliation{\jpl}
\email{eric.m.huff@jpl.nasa.gov}

\author{James G. Bartlett}
\affiliation{\jpl}
\affiliation{\cnrs}
\email{james.bartlett@jpl.nasa.gov}

\author[0000-0001-7449-4638]{Brandon S. Hensley}
\affiliation{\jpl}
\email{brandon.s.hensley@jpl.nasa.gov}

\begin{abstract}

One of the more surprising astrophysical discoveries of the last decade has been the presence of enormous quantities of dust at megaparsec distances from galaxies, which has important implications for galaxy evolution, the circumgalactic and intergalactic medium, and observational cosmology. In this work, we present a novel method for studying these vast halos of circumgalactic dust: a maximum-likelihood estimator for dust-induced extinction of background galaxies. This estimator can accommodate a broad range of archival photometric data and can incorporate different dust reddening prescriptions, making it applicable to diverse galaxy types and redshifts. We apply the estimator to the redMaGiC catalog of luminous red galaxies, selected for their tight dispersion in color and well-constrained photometric redshifts, and measure the resulting extinction as a function of projected distance from WISExSuperCOSMOS and redMaGiC foreground galaxies. We detect significant dust-induced extinction profiles extending to at least 1 megaparsec from galactic disks, with noticeable differences between star-forming and quiescent galaxies: star-forming galaxies exhibit a pronounced rise in extinction within the inner 50 kiloparsecs and a steep decline beyond 1 megaparsec, while the quiescent galaxies host little dust in the inner halo but have detectable extinction out to 30 megaparsecs. We test the robustness of our results using star catalogs and inverted foreground and background samples and find no evidence for significant systematic error. Our approach provides a powerful tool for studying the interplay between circumgalactic dust, galaxy evolution, and large-scale structure, with potential applications in a number of astrophysical subfields.

\end{abstract}




\section{Introduction}\label{sec:intro}
Most of the universe's atoms are hidden from plain sight: galaxies account for less than 10\% of baryons, while the remaining 90\% reside in gaseous phases whose low densities make them difficult to observe. These include the circumgalactic medium (CGM), which permeates the dark matter halos hosting galaxies, and the more extended intergalactic medium (IGM) that connects them \citep{2004ApJ...616..643F, 2012ApJ...759...23S}. Combined, they are the reservoir for more than 85\% of the baryons in the universe \citep{2020ApJ...902..111W}. Despite being the focus of intense theoretical and observational research, many open questions remain in the study of CGM and IGM, particularly regarding the origin, distribution, and survival of cool, condensed matter in these environments.

One pressing issue is the existence of huge amounts of dust in the CGM and IGM---comparable to the amount of dust contained within galaxies themselves---which challenges models of the cosmic baryon cycle for several reasons \citep{2020A&ARv..28....2V,2020ARA&A..58..363P,2024arXiv241210579C}.
First, dust is predominantly deposited into the \textit{interstellar} medium (ISM) through supernovae and asymptotic giant branch star winds; second, the CGM and IGM are filled with mega-Kelvin gas, which should rapidly destroy dust grains via sputtering \citep{1979ApJ...231..438D}. While a number of mechanisms for dust deposition and replenishment have been proposed, including stellar radiation-pressure driven outflows, tidal stripping of interacting galaxies, supernova-driven outflows, and cool outflows, the exact coupling between the multi-phase CGM, IGM, and ISM remains unclear \citep{2024arXiv241210579C}. 

Understanding this connection matters because the abundance and spatial extent of dust in the CGM is a key component of galaxy formation and evolution models. Dust also fundamentally limits the homogeneity of samples in wide-field surveys and introduces significant scatter to cosmological distance measurements, particularly supernovae.  According to \cite{2023arXiv231112098R}, dust is the second-largest source of uncertainty in Type Ia supernova cosmological parameter constraints, accounting for 11.5\% of the total variance. Current analyses include intergalactic dust reddening in the systematic error budget, but because the magnitude of the effect is uncertain, they do not correct the distance measures themselves. 

While an early reddening-based indication of significant quantities of dust in galaxy halos (out to at least 60 kpc from galaxy disks) appeared in \cite{1994AJ....108.1619Z}, the existence of circumgalactic dust at megaparsec scales was first reported by \citet*{2010MNRAS.405.1025M}, hereafter M10. In that work, the authors reported the serendipitous discovery of a smooth extinction profile extending to 10 Mpc while measuring the magnification of 85,000 background quasars by 20 million galaxies in SDSS images. 
A similar measurement was carried out by \cite{2015ApJ...813....7P} using a background sample of luminous red galaxies (LRGs) as ``standard crayons'' \citep{2010ApJ...719..415P}, whose small color dispersion permitted detection of reddening by dust halos with a smaller sample size than M10. Since that time, the majority of investigations of the extended CGM have been spectroscopic, e.g., \cite{2014MNRAS.439.3139Z}, \cite{2018ApJ...866...36L}, \cite{2021MNRAS.506..115Z}, and \cite{2024arXiv241108485C}. These studies typically use tracers like the equivalent width (EW) of \ion{Mg}{2} absorption lines to probe the dynamics and shape of cool gas in the CGM of galaxies at $z>0.3$, where the \ion{Mg}{2} $\lambda\lambda$ 2796, 2803 doublet is redshifted into the optical. In turn, this cool, metal-rich gas can be used to infer a dust abundance by relating the \ion{Mg}{2} absorber equivalent width to a reddening value \citep{2012ApJ...754..116M}. Far-infrared emission from CGM dust provides a more direct tracer of the dust content and thermal conditions, but current measurements are generally limited to $r \lesssim 100$ kpc from the optical radius of the galaxy \citep{2016ApJ...833..276Z, 2021MNRAS.506..115Z, 2025ApJ...990...57V}. Thus, with the notable exception of \cite{2020ApJ...905L..20R}, who reported the detection of dust-induced reddening about M31 and M33 at projected distances greater than 100 kpc, almost no studies have replicated the original M10 circumgalactic dust measurement using extinction of background sources.

These observational gaps highlight the need for a way to assess dust in the extended CGM without relying on atomic absorption lines unsuited for low-redshift galaxies, particularly given the evidence that the cosmic dust density changes significantly at $z < 1$ \citep{2020ARA&A..58..363P}.
Drawing inspiration from the ``standard crayon'' approach in \cite{2015ApJ...813....7P} and from analytical techniques ubiquitous in cosmology, we present a new technique to detect circumgalactic dust halos: a maximum-likelihood estimator for dust-induced extinction with a lower bound on the variance set by the Cram\'{e}r-Rao bound. Application of the estimator is relatively simple, requiring only a photometric dataset and specification of a theoretical model for dust-induced reddening. 
Applying this estimator to the redMaGiC galaxy catalog, notable for its tight dispersion in color, and cross-correlating the inferred extinction with $z < 0.2$ galaxies from the WISExSuperCOSMOS catalog, we find fresh evidence for circumgalactic dust halos at distances of 1 \hinv Mpc or greater from the galactic disk. We also analyze a sample of $ 0.15 < z < 0.45$ redMaGiC LRGs, and find distinct differences in the radial extinction profiles of star-forming and quiescent galaxies. 

The remainder of this paper is organized as follows. We describe our dust extinction estimator in Section \ref{sec:theory}, provide details on datasets in Section \ref{sec:catalogs}, and describe our methods in Section \ref{sec:methods}. Results are shown in Section \ref{sec:results} and discussed in Section \ref{sec:discussion}. We conclude in Section \ref{sec:conclusions}.  

\section{A maximum likelihood estimator for dust extinction}\label{sec:theory}
A screen of dust absorbs and scatters incoming light, dimming and reddening the spectrum. The extinction $A_\lambda$ quantifies wavelength-dependent dimming of light due to dust; reddening is usually quantified by a color excess, most commonly the selective extinction of red versus blue light in the UBV photometric system: $E(\mathrm{B - V})$.  
The steepness of $A_\lambda$ as a function of wavelength is parametrized by the ratio of selective-to-total extinction in V-band: $R_\mathrm{V} = A_\mathrm{V} / E(\mathrm{B - V})$. For interstellar Milky Way dust, $R_\mathrm{V} \sim 3.1$; dense molecular clouds have $R_\mathrm{V} \sim 5$ \citep{2007ApJ...663..320F}. 

Most photometric studies of the CGM trace dust using some variant of the selective extinction $E(\mathrm{B - V})$; 
here, we use maximum likelihood estimation (e.g., \citealp{barlow1993statistics}) to find a parameter that represents the best chance of finding dust-induced extinction based on a change in background galaxy colors. This estimator can be derived by maximizing the log-likelihood 
\begin{equation}\label{eqn:likelihood}
-\log L = \rm \left [\mathbf{D} - \mathbf{M}(\rho)\right ] \frac{\mathbf{C}^{-1}}{2} \left[\mathbf{D} - \mathbf{M}(\rho)\right ]^T
\end{equation}
where the data vector $\mathbf{D}$ is an array of magnitudes, e.g.,
\begin{equation}\label{eqn:data_vector}
\mathbf{D} = \{\rm g,r,i,z,...\}
\end{equation}
and the covariance of the data $\mathbf{C}$, calculated directly from the data vector $\mathbf{D}$, encapsulates both the magnitude measurement uncertainties and cosmic variance. The model $\mathbf{M}(\rho)$ describes the data $\mathbf{D}$ in terms of a parameter $\rho$ (here, $A_\mathrm{V}$); the model can be Taylor-expanded to obtain a simple linear approximation for the effect of $\rho$:
\begin{equation}\label{eqn:model}
\mathbf{M}(\rho) = \mathrm{M_0} + \rho\ \dMdp~.
\end{equation}
In this expansion, $\mathrm{M_0}$ are the intrinsic colors of galaxies in the background galaxy sample, without any reddening from dust. The $\rho$ \dMinline term is the response of galaxies to a differential change in extinction; \dMinline is a fiducial extinction curve for dust with a given $R_\mathrm{V}$. With this assumption, the optimal estimator $\hat{\rho}$ for dust-induced extinction $\rho$ such that $\rho$ maximizes the likelihood in Equation \ref{eqn:likelihood} is given as: 
\begin{equation}
\hat{\rho} = \left[ \frac{ \dMdp^T \ \mathbf{C}^{-1} \left (\mathbf{D} - \mathrm{M_0} \right )} {\dMdp^T\ \mathbf{C}^{-1 }\ \dMdp} \right ] \label{eqn:estimator}
\end{equation}
where by the Cram\'{e}r-Rao bound, the uncertainty on $\hat{\rho}$ is given by the Fischer matrix:
\begin{equation}
\sigma^2_{\hat{\rho}} = \dMdp^T\ \mathbf{C}^{-1}\ \dMdp~. \label{eqn:RCO}
\end{equation}

A potential complication to Equation \ref{eqn:estimator} is introduced by the redshift dependence of the average galaxy color $\overline{\mathrm{M_0}}$. In practice, the quantity $\hat{\rho}$ must be estimated in bins of galaxy redshift, with the exact number of bins chosen to balance the variance and change of $\overline{\mathrm{M_0}}$ as a function of redshift. 

With the extinction estimator $\hat{\rho}$ in place, foreground galaxy number counts may be cross-correlated with the dust-induced extinction of 
background galaxies. Denoting $N(\theta_1)$ as the galaxy count at a position $\theta_1$ on the sky and $A_\mathrm{V}(\theta_2)$ as the 
extinction at position $\theta_2$, the cross-correlation may be expressed as the average deviations of 
$N$ and $A_\mathrm{V}$ from their mean values at two sky positions $\theta_1$ and $\theta_2$ separated by $\theta$:
\begin{equation}
\xi_{N A_\mathrm{V}}(\theta) = \left< [N(\theta_1) - \overline{N}]\,[A_\mathrm{V}(\theta_2) - \overline{A_\mathrm{V}}] \right> \label{eqn:correl}
\end{equation}
The mean values, $\overline{N}$ and $\overline{A_\mathrm{V}}$, are the average foreground galaxy count and average extinction of background galaxies, respectively, over the entire survey area. 

As with any such analysis, the contribution of random correlations
to \av  must be taken into account. We use the following estimator for the true galaxy-reddening cross-correlation $\xi_{N A_\mathrm{V}}(\theta)$:
\begin{equation}
\xi_{N A_\mathrm{V}}(\theta) = D_N D_{A_\mathrm{V}}(\theta) - D_N R_{A_\mathrm{V}}(\theta) - R_N D_{A_\mathrm{V}}(\theta) + R_N R_{A_\mathrm{V}}(\theta)\label{eqn:xi_estimator}
\end{equation}
$D_N D_{A_\mathrm{V}}$ is the signal from the correlation of the real foreground galaxy and background color catalogs; $D_N R_{A_\mathrm{V}}$ and $R_N D_{A_\mathrm{V}}$ are the correlations of real foreground with random background catalog and random foreground with real background catalogs, respectively; and $R_N R_{A_\mathrm{V}}$ is the correlation of the random foreground and background catalogs. 


\section{Data}\label{sec:catalogs}

We use the Dark Energy Survey (DES) Y3 redMaGiC catalogs \citep{2016MNRAS.461.1431R, 2022PhRvD.106d3520P} as background samples and the WISExSuperCOSMOS photometric redshift catalog \citep{2016ApJS..225....5B}, plus a low-redshift subset of redMaGiC galaxies, as foreground samples. In addition, star catalogs selected from the Gaia survey and DES Y3 are used to test for systematic error. Further details are presented below. 
    
\subsection{Galaxy catalogs}

The redMaGiC galaxies are divided into two catalogs across five tomographic bins from $0.15 < z < 0.90$. Galaxies in the first three redshift bins ($0.15 < z < 0.65$) were drawn from a ``high density'' catalog, which is volume limited to a z-band luminosity of 0.5L*. The last two tomographic bins ($0.65 < z < 0.90$) are drawn from a ``high luminosity high redshift'' catalog, which is volume limited to a z-band luminosity of 1.0 L*. Because the galaxy population and selection function differ between the two samples, we perform our analysis on the high-density and high-redshift redMaGiC catalogs separately. An additional cut of $z > 0.5$ is placed on the high-density redMaGiC sample to ensure adequate foreground-background separation. We use the $z < 0.45$ population of the high-density redMaGiC catalog to investigate the extended CGM of luminous red galaxies.

The redMaGiC catalogs do not contain photometry and must be joined with another catalog to obtain it. 
We use CosmoHub\footnote{\url{https://cosmohub.pic.es/home}} \citep{2017ehep.confE.488C,TALLADA2020100391} to query the DES Y3 GOLD photometric catalog, using the galaxy selection criteria in \cite{2021ApJS..254...24S} to obtain a pure sample of galaxies with good photometry that can then be joined to the redMaGiC catalog.  Additional selections to exclude galaxies with NaN, $-9999$, or $< 28$ magnitudes in any bandpass are imposed; these cuts only eliminate an additional 1\% of galaxies.

The reddening calculation begins with the DES Y3 GOLD \texttt{MOF\_CM\_MAG\_}$\{$\texttt{g,r,i,z}$\}$ magnitudes. 
To correct these ``top of the Milky Way'' magnitudes for Galactic reddening, we depart from the DES Y3 default of \cite{1998ApJ...500..525S} and implement a corrected version, CSFD, which uses a large number of empirical templates to clean the canonical SFD map of contamination from large-scale structure \citep{2023ApJ...958..118C}. Specifically, we use the \texttt{dustmapper}\footnote{\url{https://github.com/gregreen/dustmaps}} package \citep{2018JOSS....3..695M} to query the CSFD dust map and obtain extinction coefficients for a given $R_\mathrm{V}$ value, set of coordinates, and set of bandpasses. 
Equation \ref{eqn:CSFD} then presents the complete expression for the corrected magnitudes $\texttt{MAG\_X\_CORRECTED\_CSFD}, \texttt{X} \in \{g,r,i,z\}$ used in our analysis. This formula replaces $A_{\rm SFD98}$ with the equivalent CSFD extinction coefficient $A_\mathrm{X,\, CSFD}$, while also incorporating updates to the photometric zero-point ($\texttt{DELTA\_MAG\_Y4\_X}$) and a correction for the spectral energy distribution of calibration sources ($\texttt{DELTA\_MAG\_CHROM\_X}$), following the prescription of \cite{2022PhRvD.106d3520P}. We evaluate the effect of the SFD98 extinction prescription in Appendix~\ref{sec:csfdvsnot}. 
\begin{align}\label{eqn:CSFD}
\begin{split}
\texttt{MAG\_X\_CORRECTED\_CSFD} &={} 
	\texttt{MOF\_CM\_MAG\_X} + \texttt{DELTA\_MAG\_Y4\_X} \\
	& + \texttt{DELTA\_MAG\_CHROM\_X} - \texttt{A\_X\_CSFD}
\end{split}
\end{align}

\begin{figure*}[hbtp]
    \begin{center}
    \includegraphics[width=0.95\linewidth]{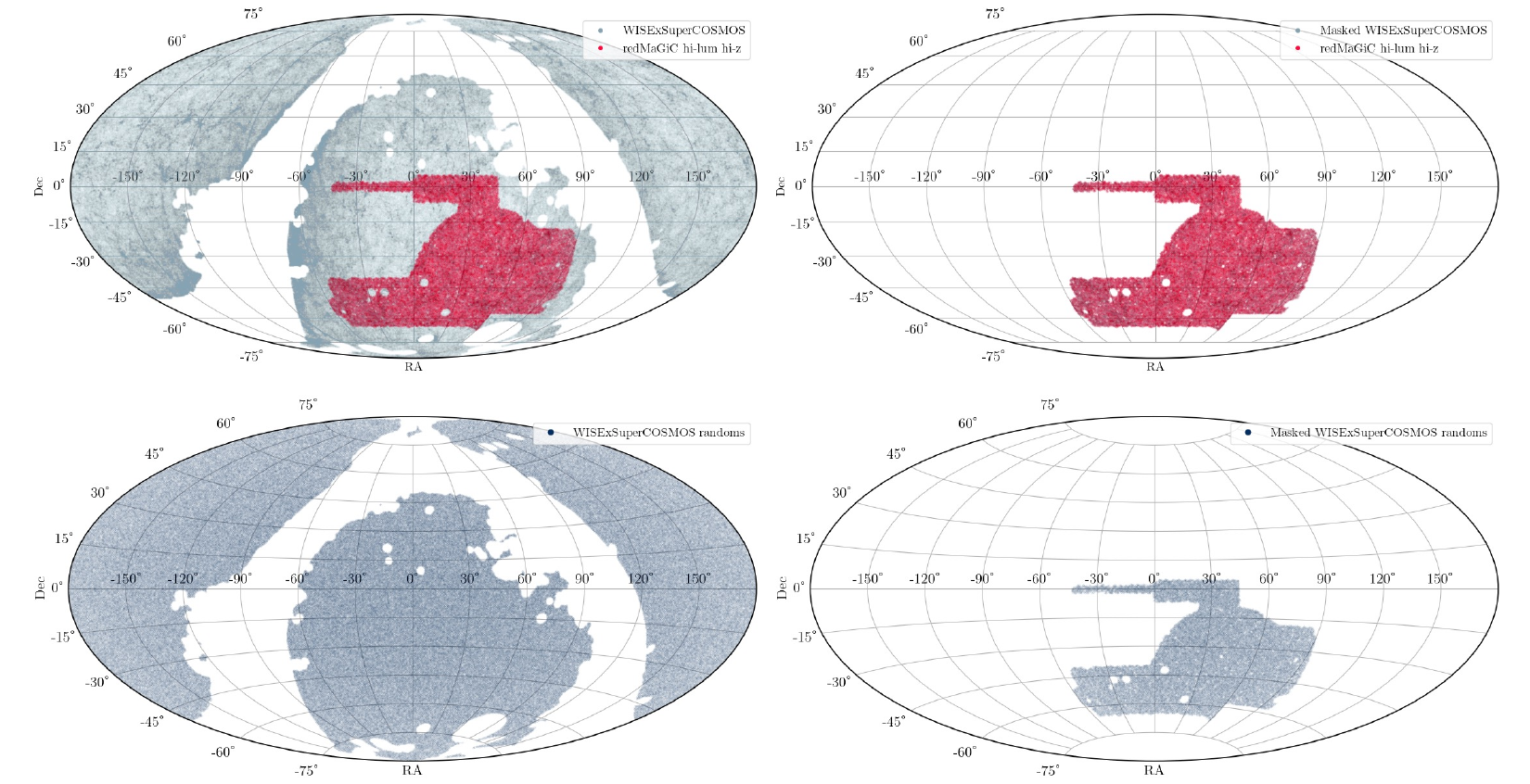}
    \caption{Sky coverage of catalogs used in this analysis. Top row: foreground WISExSCOS (grey points) and the background high-luminosity high-redshift redMaGiC galaxy catalog (fuchsia points) before (left panel) and after (right panel) masking. Bottom row: WISExSCOS random galaxy catalog after application of the WISExSCOS survey mask alone (left panel) and after application of both WISExSCOS and DES survey masks (right panel).}
    \label{fig:galaxy_overlap}
    \end{center}
\end{figure*}

The WISExSuperCOSMOS\footnote{\url{http://ssa.roe.ac.uk/WISExSCOS.html}} (or WISExSCOS) catalog combines mid-infrared data from the Wide-field Infrared Survey Explorer (WISE) and optical data from SuperCOSMOS scans of UKST/POSS-II photographic plates \citep{2016ApJS..225....5B}. The full catalog contains 20 million sources and covers the whole sky; our analysis is restricted to the subset of galaxies that overlap with the DES Y3 field and have corrected photometric redshifts $0< z < 0.2$. An additional cut of $B^\mathrm{cal}_\mathrm{corr} > 16 $ is made to filter out stellar contaminants and very large galaxies that might be blended with background galaxies. Sky coverage of the WISExSCOS catalog, as well as redMaGiC, is shown in the top row of Figure~\ref{fig:galaxy_overlap}. 

Random (mock) galaxy catalogs, needed for the calculation of $\xi_{NA_\mathrm{V}}(\theta)$, are provided for the two redMaGiC galaxy samples but not for WISExSCOS. Accordingly, we generate a random catalog for WISExSCOS by first sampling a large number of points uniformly in $(\theta,\sin\phi)$ then applying the WISExSCOS survey mask. 
The result is shown in the bottom left panel of Figure \ref{fig:galaxy_overlap}. Additional application of the DES Y3 survey mask produces the map shown in the bottom right panel of Figure~\ref{fig:galaxy_overlap}.  

The final catalog sizes are: 1,293,547 galaxies in the WISExSuperCOSMOS catalog; 871,556 galaxies at $z > 0.5$ in the high-density redMaGiC background catalog; 34,711,359 objects in the $z > 0.5$ random high-density redMaGiC catalog; 816,204 galaxies in the high-luminosity high-redshift redMaGiC catalog; 32,597,284 objects in the random high-luminosity high-redshift redMaGiC catalog; 660,074 galaxies in the $z < 0.45$ high-density redMaGiC foreground catalog; and 26,306,402 objects in the random $z < 0.45$ redMaGiC catalog. 

\subsection{Star catalogs for systematics testing}

Foreground star catalogs (for cross-correlation with the redMaGiC background catalogs) 
are selected from the Gaia Data Release 3 database \citep{2023A&A...674A...1G}, accessed via CosmoHub.
There is a strong gradient in the surface density of Gaia stars towards the galactic plane; failing to account for this variation in source density in random catalogs might bias the resulting correlation measurement. Accordingly, both ``signal'' and ``random'' star catalogs are independent, random subsamples ($\sim$ 478,000 objects) of the full Gaia Data Release 3 catalog. As elsewhere, the region of overlap with DES is obtained by calculating the intersection of survey masks for Gaia and DES. 

To create a background star catalog (for cross-correlation with the WISExSuperCOSMOS foreground), we query the DES Y3 GOLD photometric catalog using the star selection criteria in \cite{2021ApJS..254...24S}. An additional cut of $18 <$ \texttt{mag\_r\_corr} $< 21$ eliminates saturated stars and possible galaxy confusion. The final catalog sizes are 824,345 stars in the ``signal'' catalog and 9,892,452 stars in the ``random'' catalog. Because these star catalogs contain the same magnitude measures as the joined redMaGiC-Y3 GOLD galaxy catalog, comparison of results between the two is very straightforward.

\section{Methods}\label{sec:methods}

This section presents the analysis pipeline we put in place to apply the formalism of Section \ref{sec:theory} and so measure circumgalactic dust extinction profiles. The major steps of the pipeline are: intake and processing of foreground and background catalogs including random catalogs; defining a particular dust extinction model; dividing of galaxies into redshift bins and de-meaning of galaxy colors; per-galaxy calculation of the value and uncertainty of \av; calculation of $\xi_{N A_\mathrm{V}}(\theta)$; 
and finally, the fitting of a polynomial to the resulting extinction profile. Details are provided in the subsections below. The code developed to carry out this analysis, \texttt{dusthalos}, is hosted on GitHub\footnote{\url{https://github.com/mcclearyj/dusthalos/}} and will be made publicly available upon acceptance of this manuscript. 

\subsection{Catalog handling}\label{sec:catalog_handling}

The high-level \texttt{dusthalos} workflow for input galaxy catalog handling is: 

 \begin{enumerate}
    \item Load in the foreground and background catalogs and their 
    respective survey masks. If a mask is not available for a given survey, create one.
    \item Apply survey masks to the catalogs, only retaining galaxies that 
    lie in the intersection of both. 
    \item If random (mock) galaxy catalogs are not available for a given survey, create one. 
    \item Correct background galaxy magnitudes for Milky Way extinction. If the background galaxy catalog does not include photometric information, join it to a photometric catalog. 
    \item Save masked and augmented catalogs to file.
 \end{enumerate}

The specific implementation of this workflow for the galaxy and star catalogs used in our analysis is mostly covered in Section \ref{sec:catalogs}, with a few additional details below. 

\textit{Masking.} Although HEALPix survey masks are available for the DES Y3 and WISExSuperCOSMOS catalogs, they are not available for Gaia. To ensure an exact match in sky coverage of the foreground and background catalogs, we create our own using coordinate information in the catalog and the healpy Python package \citep{Zonca2019}. 

 
\begin{figure}[hbt]
   \centering
   \includegraphics[width=0.8\linewidth]{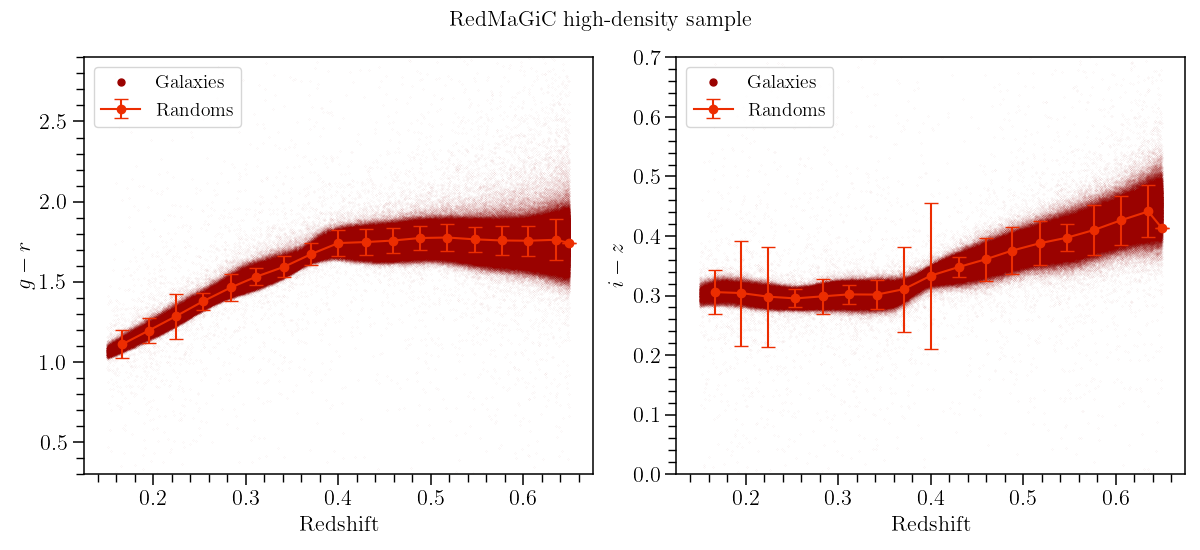}\\
   \includegraphics[width=0.8\linewidth]{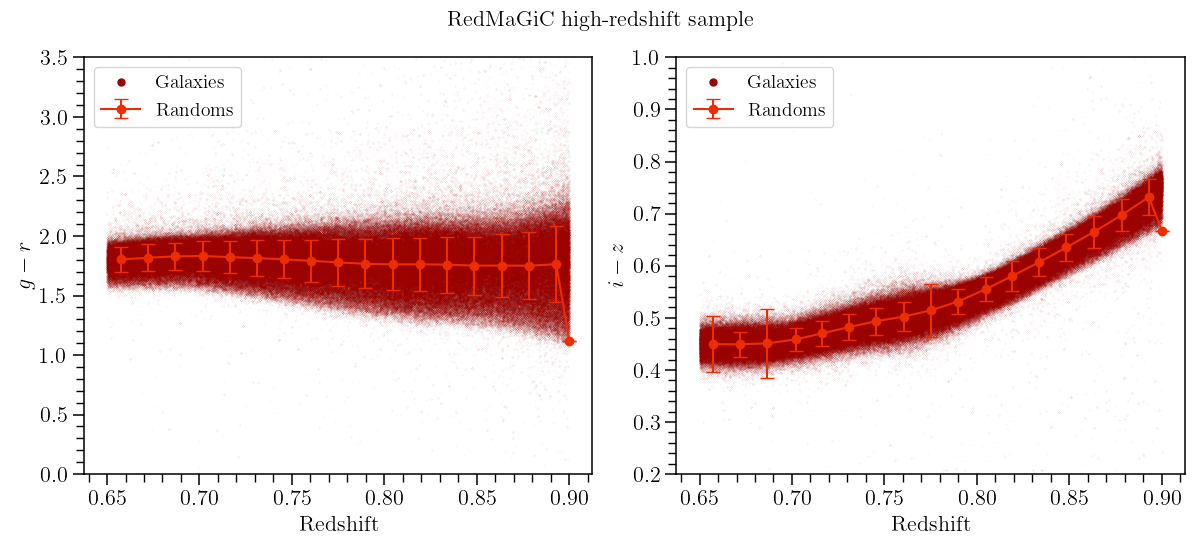}
   \caption{Color versus redshift for the two redMaGiC galaxy catalogs (small, dark red points) and their respective random galaxy catalogs (large, bright red points connected by a solid line). Top row: high density galaxy sample. Bottom row: high-luminosity high-redshift sample.}
   \label{fig:redshift_match}
\end{figure}

\textit{Photometry.} The redMaGiC catalogs we use in our analysis include celestial coordinates and redshifts but not magnitudes. As described in Section \ref{sec:catalogs}, to obtain colors for the real redMaGiC catalogs, we join member galaxies to their counterparts in the Y3 GOLD catalog on the \texttt{COADD\_OBJECT\_ID} parameter. Random catalog entries lack a corresponding entry in the Y3 GOLD catalog and so are randomly assigned photometry from the appropriate redshift slice in the photometry-augmented redMaGiC galaxy catalog. The color-matched random catalogs resulting from this procedure are shown in Figure \ref{fig:redshift_match}, which shows good agreement with the real data. 

\subsection{Extinction calculation and correlation}\label{sec:extinction_code}


To obtain values for \dMinline, the model for dust-grain induced reddening referenced in Equations \ref{eqn:model}--\ref{eqn:RCO}, we use the comprehensive dust extinction law published by \cite{2023ApJ...950...86G}, hereafter G23. 
The G23 model requires specification of the wavelengths at which to compute the extinction and a value for $R_\mathrm{V}$. For the former, we use the effective wavelengths of the DECam filter set available through the Virtual Observatory's Filter Profile Service, rescaled as prescribed in Table 6 of \cite{2011ApJ...737..103S}: $\left\{g, r, i, z \right\} = \left\{4796.6, 6382.6, 7769.0, 9108.2\right\}$. For the latter, in the absence of a compelling reason to do otherwise, we adopt the common default of $R_\mathrm{V}= 3.1$. (Further consideration is given to the choice of $R_\mathrm{V}$ in Section \ref{sec:discussion}.) 

After selection of \dMinline, galaxies are sorted into redshift bins, and within each bin, the weighted mean galaxy color is computed using individual galaxy magnitude uncertainties as weights. The per-bin mean is then subtracted from the individual galaxy magnitudes. For this calculation, we adopt the default redMaGiC redshift binning scheme: three for the high-density sample, two for the high-luminosity high-redshift sample. We then compute values of the optimal estimator for extinction (Equation \ref{eqn:estimator}) and the corresponding minimum bound on the uncertainty (Equation \ref{eqn:RCO}) for each galaxy, obtaining the inverse covariance matrix $\mathbf{C}^{-1}$ directly from the galaxy magnitudes. This operation is repeated for both real and random background catalogs. 

We compute $\xi_{N A_\mathrm{V}}(\theta)$ using the galaxy-kappa cross-correlation method \texttt{NKCorrelation} provided in the TreeCorr package\footnote{\url{https://github.com/rmjarvis/TreeCorr}} \citep{2004MNRAS.352..338J}. Though intended to correlate galaxy counts with convergence (i.e., $\kappa$) the code is flexible enough to accommodate any scalar quantity such as $A_\mathrm{V}$.  
All correlations are taken in 13 impact parameter bins, from 0.5 to 200 arcminutes, or about 20 kiloparsecs to 10 megaparsecs in the average foreground galaxy rest frame. This range matches the physical scales probed in M10 and related works.
Error bars on extinction profiles are based on TreeCorr's \texttt{sample} covariance estimation method, which is the simplest method available in that package to account for the combination of Poisson counting statistics, statistical variance of the scalar quantity, and sample variance across the observation. Finally, to facilitate comparison across galaxy samples and the literature, we finally fit power laws of the form $A_\mathrm{V}(r) = C r^\alpha$ to the extinction profiles using weighted least squares. 

Our pipeline enables several tests for systematic error. Substitution of star catalogs for galaxy foregrounds and backgrounds offers a valuable null test for our pipeline, as main-sequence stars do not have extended halos of dust. Similarly, stars embedded in the Milky Way should not display dust-induced color gradients as a function of projected distance from galaxies outside the Milky Way. 
Additionally, inverting foreground and background samples, i.e., using a high-redshift catalog as a foreground and a low-redshift catalog as a background, should result in a non-detection of reddening. We deploy all three of these tests in the course of our analysis. Additional systematic errors that could affect our results are considered in Appendix \ref{sec:other-systematics}. 

\section{Results}\label{sec:results}
We present our main result, a set of circumgalactic dust halo extinction profiles, in Figure \ref{fig:redmagicfits}. For comparison, we also overplot the M10 power law (Equation 13 in that reference, reproduced below as Equation \ref{eqn:menard}):
\begin{equation}
A^\mathrm{M10}_\mathrm{V}(r) = \left(4.4 \pm 1.1\right) \times 10^{-3} \left( \frac{r}{100 h^{-1}\, \mathrm{kpc}} \right)^{-0.86 \pm 0.19}\label{eqn:menard}
\end{equation}
%
Best-fit power laws to our own extinction profiles are shown in Equations \ref{eqn:stacked} and \ref{eqn:lrg}. The outcomes of systematics tests are summarized in Table \ref{tab:systematics} and Figure \ref{fig:systematics}. More detail is provided in the subsections below. For the interested reader, we also compare extinction profiles obtained with the CSFD galactic extinction correction to profiles obtained with the canonical SFD extinction correction in Appendix~\ref{sec:csfdvsnot}.
 
\subsection{Galaxy extinction profiles}\label{sec:redmagic_results}

\begin{figure*}[htb]
\begin{center}
\includegraphics[width=0.75\textwidth]{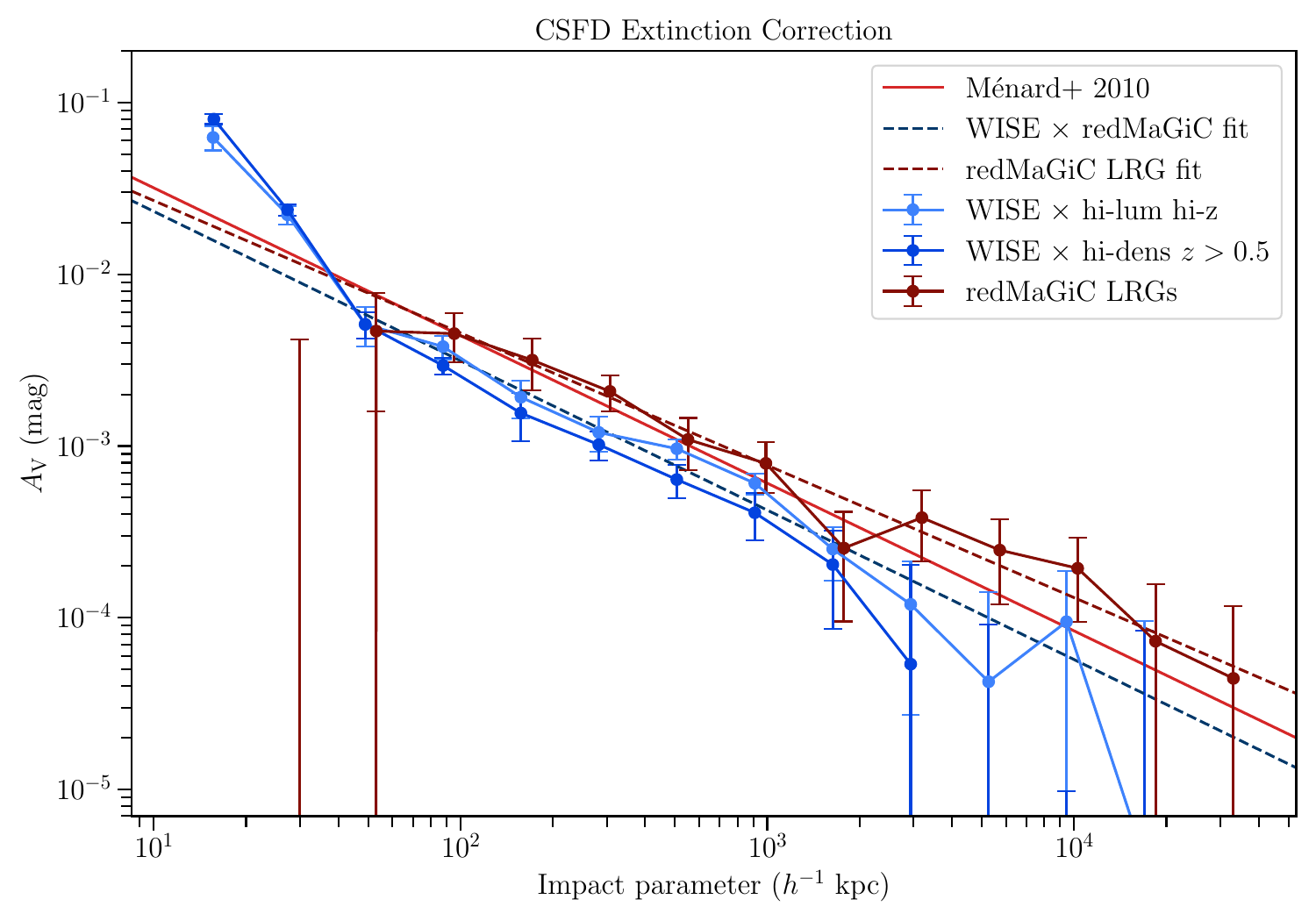}
\caption{Extinction vs impact parameter for WISExSuperCOSMOS (blue points) and redMaGiC LRGs (maroon points). Power-law fits are shown for the redMaGiC LRG profile (maroon dashed line) and the combined WISExSuperCOSMOS x redMaGiC profiles (blue dashed line). For comparison, the M10 result is shown in red. Results are based on the CSFD Milky Way extinction correction.}
\label{fig:redmagicfits}
\end{center}
\end{figure*}

The extinction profiles of WISExSuperCOSMOS galaxies are shown in Figure \ref{fig:redmagicfits}: the result from cross-correlation with the $z > 0.5$ high-density redMaGiC sample is plotted in dark blue, and the result from cross-correlation with the high-redshift redMaGiC sample is plotted in light blue. Both extinction profiles follow a broken power law, with a rapid decrease from $A_\mathrm{V} \sim 0.3$ to $A_\mathrm{V} \sim 0.02$ in the inner 50 \hinv kpc, then a slower decline to $A_\mathrm{V} \sim 0.02$ at 1 \hinv Mpc. At the transition to the 2-halo regime ($r \gtrsim 2$ \hinv Mpc), the extinction profiles rapidly diminish to below our detection threshold ($A_\mathrm{V} \sim 10^{-5}$). 

We fit a power law to the WISExSuperCOSMOS extinction profiles in the 1-halo regime (50 \hinv kpc $ < r < 1$ \hinv Mpc), where the lower limit is set to the point where the redMaGiC LRG and WISExSuperCOSMOS extinction profiles intersect, and the upper limit is set to the point where the extinction profiles taper off. Although the selection functions of the two redMaGiC background catalogs are different, the foreground catalog is of course the same. Accordingly, we fit the two WISExSuperCOSMOS $\times$ redMaGiC profiles with a single power law, which yields  Equation \ref{eqn:stacked}. Quoted uncertainties are the standard error on the mean parameter values. 
%
%
\begin{equation}
A_{\mathrm{V,\,stacked}}(r) = \left(3.1\pm 0.06\right) \times 10^{-3} \left( \frac{r}{100\, h^{-1}\, {\rm kpc}} \right)^{-0.87 \pm 0.04}\label{eqn:stacked}
\end{equation}
\noindent

Cross-correlation of the redMaGiC LRG galaxy sample with the redMaGiC high-luminosity high-$z$ sample yields the extinction profile plotted in maroon in Figure \ref{fig:redmagicfits}. As might be expected for this galaxy population, there is little apparent dust at the lowest impact parameters, within a few optical radii of the galaxies ($r < 50$ \hinv kpc). Between $0.05 < r < 2$ \hinv Mpc, the LRG profile is qualitatively similar to the WISExSCOS profiles. However, the LRG sample shows significant extinction out to 30 \hinv Mpc, with a distinct dip at $\sim 2$ \hinv Mpc---the point at which the WISExSCOS profiles taper off. We fit a power law to the LRG \av profile in the range 50 \hinv kpc $ < r < 30$ \hinv Mpc, with the result given in Equation \ref{eqn:lrg}. As before, quoted uncertainties reflect the standard error on the mean parameter values. 
\begin{equation}
A_{\mathrm{V, \, LRG}}(r) = \left(4.6 \pm 0.2\right) \times 10^{-3} \left( \frac{r}{100\, h^{-1}\, \mathrm{kpc}} \right)^{-0.77 \pm 0.04}\label{eqn:lrg}
\end{equation}
%

\subsection{Systematics tests}\label{sec:systematics_results}

\begin{figure*}[htb]
\begin{center}
\includegraphics[width=0.85\linewidth]{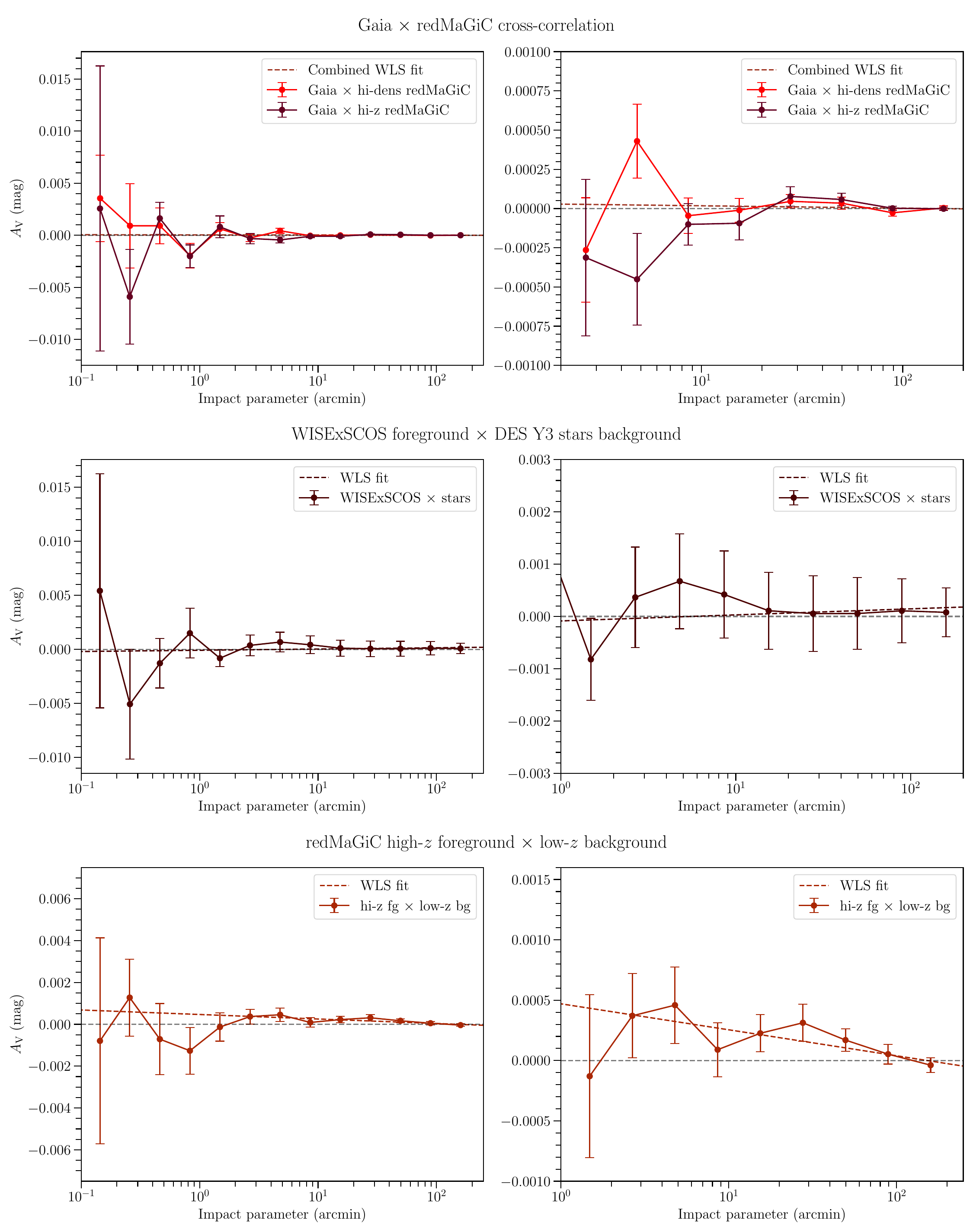}
\caption{Extinction profiles produced for systematics testing with best-fit power laws overplotted. Left panels show the full impact parameter range; right panels are cropped to high impact parameters. Top row: Gaia foreground cross-correlated with redMaGiC high-$z$ (maroon) and high-density (red) background catalogs. 
Middle row: WISExSuperCOSMOS galaxy foreground cross-correlated with DES Y3 stars. Bottom row: cross-correlation of redMaGiC high-$z$ foreground with $z<0.45$ high-density background.} 
\label{fig:systematics}
\end{center}
\end{figure*}

Figure~\ref{fig:systematics} presents the results of the systematics tests described in Section \ref{sec:methods}, namely, the calculation of extinction profiles using star catalogs and with inverted foreground and background samples (high-$z$ foreground, low-$z$ background). The top row of Figure~\ref{fig:systematics} shows the the extinction profile resulting from the cross-correlation of the Gaia star foreground catalog with the redMaGiC high-luminosity high-redshift sample (maroon points) and high-density sample (red points). The middle row of Figure~\ref{fig:systematics} shows the \av profile resulting from the cross-correlation of the WISExSuperCOSMOS foreground with the background catalog created from DES Y3 stars. 
For the reverse \av profile test, we invert the redMaGiC LRG measurement and cross-correlate the high-redshift redMaGiC catalog with the $z<0.45$ high-density redMaGiC catalog. The outcome is shown in the bottom row of Figure \ref{fig:systematics}. 

As with the galaxy samples, we fit power laws to the systematics-test extinction profiles, including the combined Gaia-redMaGiC high-density and high-redshift profiles. The resulting extinction profile parameters are summarized in Table \ref{tab:systematics} and also plotted in Figure~\ref{fig:systematics}. 
Weighted least squares fits to these extinction profiles are effectively all consistent with zero, with prefactors close to unity and exponents on the order of $10^{-5}$. The only fit that yielded a moderately significant ($3\, \sigma$) exponent was for the inverted foreground-background \av profile. Even so, the magnitude of this exponent is only one ten-thousandth of the magnitude of the galaxy \av profile exponents, indicating no cause for concern.

\begin{deluxetable}{ccc}\label{tab:systematics}
\tablecaption{Systematics test \av profile fits}
\tablehead{
\colhead{Sample} & \colhead{Prefactor} & \colhead{Exponent ($\times 10^{-5}$)}
}
\startdata
Gaia \x redMaGiC high-density & $1.0\, \pm \, (8.0 \ee{-5})$ &  
$-1.0 \, \pm$  1.7\\
Gaia \x redMaGiC high-$z$ & $1.0\, \pm \, (9.7 \ee{-5})$ & 
$-0.3 \, \pm \, 2.0$\\
Gaia \x redMaGiC combined & $1.0\, \pm \, (6.0 \ee{-5}$) &
$-0.7 \pm 1.3$\\
WISExSCOS \x DES stars & $1.0 \, \pm \, (3.0 \ee{-4}$) & 
$5.0\, \pm \, 8.3$ \\
redMaGiC high-$z$ \x low-$z$ & $1.0 \, \pm \, (1.3 \ee{-4}$) & 
$-9.4\, \pm \, 3.0$ \\
\enddata
\end{deluxetable}

%

\section{Discussion}\label{sec:discussion}
Figure \ref{fig:redmagicfits} shows an unambiguous detection of $A_\mathrm{V}$ at megaparsec scales about both redMaGiC LRGs and a mixed sample of quiescent and star-forming galaxies in WISExSCOS, with negligible contributions from systematic errors (Figure \ref{fig:systematics} and Table \ref{tab:systematics}). The slopes ($\alpha \simeq {0.8}$) of the extinction profiles presented in Equations \ref{eqn:stacked} and \ref{eqn:lrg} are in excellent agreement with M10 (cf.~Equation \ref{eqn:menard}) and other studies both EW(\mgII)-based  \citep{2014MNRAS.439.3139Z,2018ApJ...866...36L,2024arXiv241108485C} and extinction-based \citep{2020ApJ...905L..20R}. This ubiquitous value of $\alpha \simeq 0.8$ is directly connected to halo mass profiles, where the so-called differential surface density typically follows $\Delta\Sigma\propto r^\beta$. For massive, dynamically relaxed galaxies, $0.7 \le \beta \le 0.9$ \citep{2013MNRAS.432.1046B,2014MNRAS.437.2111V,2016MNRAS.459.3251V}. 
These values of $\beta$ match both our measured $\alpha \simeq 0.8$ and values of $\alpha$ in the literature, which implies that dust permeates the entire galaxy halo. 
In addition to the exponent, the normalization of our $A_{\mathrm{V, \, LRG}}$ is also in excellent agreement with M10, while the $A_{\mathrm{V, \, stacked}}$ is about $1 \, \sigma$ lower. 
Although a direct comparison of our result to \mgII absorbers is impossible without spectroscopic calibration of our samples, by applying the \cite{2008MNRAS.385.1053M} scaling between $E(B-V)$ and EW(\mgII), we find a (very) approximate EW of 1.4 \AA~at an impact parameter of 20 kpc for our WISExSCOS sample, within a factor of 2 of the ELG results of \cite{2018ApJ...866...36L} and \cite{2024arXiv241108485C}. 

We may convert our observed $A_\mathrm{V}$ profiles into an approximate dust mass surface density profile $\Sigma_d(r)$ using the astrodust+PAH model of Milky Way dust described in \cite{2023ApJ...948...55H}. Using the astrodust surface dust mass $\Sigma_d = 1.18\ee{-26} \, \mathrm{g} \, N_H$ and adopting the extinction per hydrogen column density  $A_\mathrm{V}/N_H = 3.24\ee{-22}$ mag cm$^2$ from Table 2 of \cite{2023ApJ...948...55H}, we find $\Sigma_d/A_\mathrm{V} = 0.2\,{\rm M}_\odot\, \mathrm{pc}^{-2}\, \mathrm{mag}^{-1}$. This estimate assumes that CGM dust has the same physical properties, e.g., composition and size distribution, as dust in the local ISM, an assumption which is unlikely to be accurate in detail (see discussion below). More refined constraints on the inferred dust mass would be possible if the full wavelength dependence of the extinction were to be determined in the future.

There are some notable differences between the WISExSuperCOSMOS and LRG \av profiles, the most obvious of which is the sharp rise in the WISExSuperCOSMOS extinction at $r < 50\ h^{-1}\, \mathrm{kpc}$.
This excess is consistent with the inclusion of star-forming galaxies with dense inner CGM in the WISExSuperCOSMOS sample.  
Conversely, the LRG extinction profile has essentially no dust at these radii, which is consistent with LRGs having had no recent episodes of star formation. 
Although an inner \av excess was not evident in the original M10 result, numerous studies have since confirmed the broken power law shape of dust-induced absorption about star-forming (emission line) galaxies, \textit{inter alia}, \cite{2018ApJ...866...36L,2020ApJ...905L..20R, 2024arXiv241108485C}. 
In fact, the WISExSuperCOSMOS profiles of Figure \ref{fig:redmagicfits} bear a striking resemblance to Figure 4 of \cite{2024arXiv241108485C}, in which the EW(\mgII) profiles of ELGs and LRGs and are superimposed and which features a very similar transition from high to low absorption at $\sim\!60$ \hinv kpc. 

At larger angular scales, the amplitude of the LRG extinction profile is systematically higher than the WISExSuperCOSMOS sample and remains detectable out to 30 Mpc, with a slight dip in \av at $R \approx 2 \, h^{-1}\, \mathrm{Mpc}$, near the boundary between the one-halo and two-halo regimes. Meanwhile, the WISExSuperCOSMOS \av tapers off at $R \gtrsim 1 \, h^{-1}$ Mpc. The difference in \av alone is difficult to interpret, since the density of dust in the CGM depends on stellar mass, redshift, and star formation rate. Moreover, the redMaGiC LRG sample ($\bar{z} = 0.35$) spans a larger cosmological volume than the WISExSCOS sample ($\bar{z} = 0.14$), implying that the LRG sample's dust column density should also be larger. 
Even so, a few qualitative statements can be made. LRGs tend to reside in denser environments where interactions are more frequent \citep{1984ApJ...281...95P,2006MNRAS.373..469B,2024ApJ...962...58C}, offering more mechanisms for dust deposition and shielding (e.g., \citealp{boselli2022ram}). By contrast, star-forming galaxies such as those in the WISExSCOS sample are on average lower-mass and less likely to reside in dense environments, thus dispersing their dust into a smaller volume within a relatively empty IGM where the dust is ultimately destroyed. Indeed, when comparing Figure \ref{fig:redmagicfits} with the \mgII surface mass profiles in Figure 5 of \cite{2014MNRAS.439.3139Z}, our $A_{V, \,\mathrm{stacked}}$ resembles a one-halo profile, while $A_{V, \,\mathrm{LRG}}$ appears to be the superposition of one-halo and two-halo components. 

The transition at 2 \hinv Mpc in both the LRG and WISExSCOS dust-extinction profiles is particularly interesting in the context of the ``tired wind'' model for galactic outflows driven by star formation \citep{2018MNRAS.481.1873L, 2021MNRAS.503..336K}. 
In this framework, galactic wind-driven bubbles---hot, fast outflows powered by supernova energy injection---expand and cool over Gyr timescales, ultimately ``hanging'' in the outer halo. While the \cite{2018MNRAS.481.1873L}  model tracks the evolution of a single wind bubble for an individual galaxy---a good description of the high \av in the inner halo of star-forming WISExSCOS galaxies---the observed LRG extinction profile can be interpreted as the collective effect of many such bubbles, launched at different epochs by different galaxies, persisting in the CGM over cosmic timescales. 
Based on the theoretical expectation that bubble sizes grow slowly to hundreds of kiloparsecs and that entrained dust can survive for Gyr \citep{2024ApJ...974...81R}, it is physically consistent for LRGs to host ancient, radially outflowing wind bubbles at 2–3 Mpc. This is also close to the spike in \mgII absorber redshift space distortions and truncation radius shown in \cite{2021MNRAS.506..115Z}. 

Finally, we consider the implications of our choice of dust model. All results presented in this work necessarily assume a form for the dust extinction law \dMinline shown in Equations~\ref{eqn:estimator} and \ref{eqn:RCO}. In turn, \dMinline necessarily assumes some physical characteristics for dust that govern the wavelength dependence of the extinction. Our adoption of G23 with $R_\mathrm{V} = 3.1$ is tantamount to stating that dust in the CGM is the same as dust in the Milky Way ISM. In reality, dust composition likely differs between the two environments: large dust grains are more likely to survive transport from the ISM to the CGM in hot winds; meanwhile, small grains are likely formed \textit{in situ} via shattering of larger grains \citep{2024PASJ...76..753H, 2024ApJ...974...81R}. Although clearly not a perfect representation of the CGM, use of a Milky-Way type \dMinline model at least allows for easy comparison with the literature. We plan to evaluate the impact of different dust models on observed \av profiles in future work. 

\section{Conclusions and Future Directions}\label{sec:conclusions}

In this work, we introduce a new technique for studying circumgalactic dust halos: a maximum-likelihood estimator for dust-induced extinction. Our implementation features up-to-date models for dust-induced reddening \citep{2023ApJ...950...86G} and foreground Galactic extinction \citep{2023ApJ...958..118C}. We apply our estimator to two redMaGiC galaxy catalogs with distinct selection functions and perform angular cross-correlation of the resulting extinction with two foreground galaxy samples: a $z < 0.2$ subset of WISExSuperCOSMOS galaxies and redMaGiC LRGs at $0.15 < z < 0.45$. We detect extinction profiles consistent with megaparsec-scale dust halos in both galaxy samples, with several differences between the two. While the LRG sample displays no extinction in the inner halo ($r \leq 50$ \hinv kpc), the WISExSCOS sample, which includes star-forming galaxies, exhibits a pronounced rise in its extinction profile. At larger scales, the LRG extinction signal remains systematically higher out to 30 \hinv Mpc, with a noticeable dip at 2 \hinv Mpc, while the WISExSCOS profile tapers off beyond 1--2 \hinv Mpc. These patterns are broadly consistent with previous observations of CGM dust and mirror the expected star-forming versus quiescent galaxy populations of the two samples. We perform several tests for systematic error in our pipeline, including substituting star catalogs for both foreground and background catalogs and inverting the redMaGiC LRG samples. These tests indicate no significant biases in our measured extinction profiles, reinforcing the robustness of our main findings.


While spectroscopic investigations offer invaluable information about kinematics of the CGM, our technique offers several advantages, including suitability for galaxies at $z \lesssim 0.2$ compared to the $z \gtrsim 0.3$ required for EW(\ion{Mg}{2}) studies. Since the estimator can accommodate any combination of wavelengths, it can be applied to any archival photometric catalog. Our estimator can also accommodate a wide range of dust reddening models. 

Overall, our analysis offers a promising starting point for deeper investigations into the interplay between circumgalactic dust, galaxy properties, and large-scale structure. In future work, we plan to study the properties of CGM dust halos as a function of galaxy star formation rate, mass, and redshift, and explore the impact of different dust models on the shape and extent of extinction profiles. We will also explore the potential to derive improved calibrations for foreground reddening in supernova surveys.

\begin{acknowledgments}\label{sec:acknowledgments}
The authors thank Bryanne McDonough for insightful discussions on the topic of CGM dust and helpful feedback on the manuscript. 
Support for this work was provided by NASA through ADAP grant 18-2ADAP18-0187 (``STARDUST: Searching The hAlo Regions for Dust''). 
This work has made use of CosmoHub, developed by PIC (maintained by IFAE and CIEMAT) in collaboration with ICE-CSIC. CosmoHub has received funding from the Spanish government (grant EQC2021-007479-P funded by MCIN/AEI/10.13039/501100011033), the EU NextGeneration/PRTR (PRTR-C17.I1), and the Generalitat de Catalunya. This research has made use of the SVO Filter Profile Service ``Carlos Rodrigo", funded by MCIN/AEI/10.13039/501100011033/ through grant PID2020-112949GB-I00.
\end{acknowledgments}

%


\software{Astropy \citep{2013A&A...558A..33A,2018AJ....156..123A},
NumPy \citep{harris2020array},
TreeCorr \citep{2004MNRAS.352..338J}, 
healpy \citep{Zonca2019}
}\label{sec:software}



\appendix

\section{Comparison between SFD and CSFD}\label{sec:csfdvsnot}

\begin{figure*}[htb]
\begin{center}
\includegraphics[width=\textwidth]{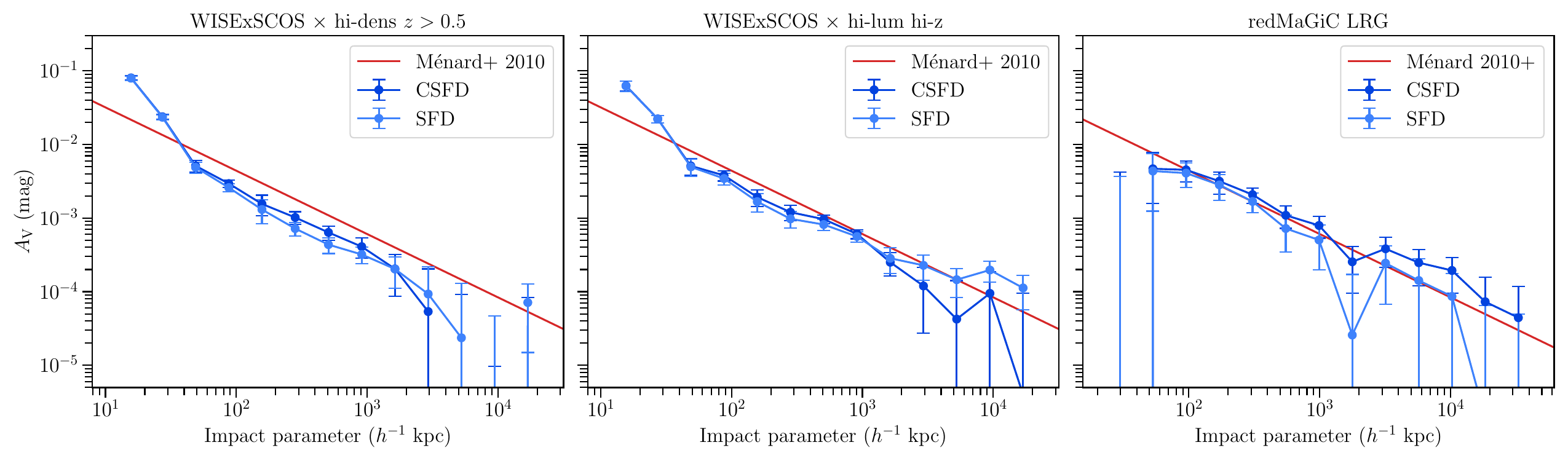}
\caption{Comparison of extinction profiles obtained with the SFD (light blue) and CSFD (medium blue) Galactic extinction corrections. From left to right, profiles are shown for WISExSuperCOSMOS galaxies cross-correlated with high-density redMaGiC background galaxies, for WISExSuperCOSMOS galaxies cross-correlated with high-luminosity high-redshift redMaGiC background galaxies, and for high-density redMaGiC galaxies cross-correlated with high-luminosity high-redshift redMaGiC background galaxies. The M10 profile is plotted in red.}
\label{fig:csfdvsnot}
\end{center}
\end{figure*}

Given the acknowledged problems of SFD \citep{10.1093/pasj/59.1.205, 10.1093/pasj/65.2.43, 2017ApJ...846...38L, 2019ApJ...870..120C}, the CSFD formulation is the default Milky Way extinction correction used in this work. It may nevertheless be of interest to compare the profiles obtained with the SFD and the CSFD prescriptions, which is shown in Figure \ref{fig:csfdvsnot}. While implementing the SFD extinction correction did not significantly alter the profile of the high-density redMaGiC background catalog, it noticeably decreased the amplitude of the high-luminosity high-redshift redMaGiC profile at large angular separations. Put differently, the SFD prescription seems to undercorrect Galactic extinction in the high-redshift catalog profile at high impact parameters, but with CSFD, the two redMaGiC \av profiles are more consistent. On the other hand, compared to the SFD prescription, the CSFD prescription results in a significant amount of extinction at high impact parameters for the redMaGIC LRG sample, past 1 \hinv Mpc. Although quite different from the effect for the WISExSuperCOSMOS result, this behavior for LRG is consistent with the idea that SFD overcorrected Galactic extinction due to contamination from the LSS. If one assumes that LRGs are a more biased tracer of the LSS than the average WISExSuperCOSMOS galaxy, then it follows that the CSFD prescription adds extinction back in. 

For the sake of comparison, we also fit power laws to the SFD-based extinction profiles.
The best-fit power laws for the WISExSCOS foreground obtained with the high-density, high-redshift, and stacked redMaGiC profiles are given by Equations \ref{eqn:hidens_notcsfd}, \ref{eqn:hiz_notcsfd}, and \ref{eqn:stacked_notcsfd}, respectively: 
\begin{align}
A_{\mathrm{V,\,hidens}}(r) &= \left(2.0\pm 0.2\right) \times 10^{-3} \left( \frac{r}{100\,  h^{-1}\, \mathrm{kpc}} \right)^{-0.85 \pm 0.02}\label{eqn:hidens_notcsfd}\\
A_{\mathrm{V,\,hiz}}(r) &= \left(2.6 \pm 0.3\right) \times 10^{-3} \left( \frac{r}{100\, h^{-1}\, \mathrm{kpc}} \right)^{-0.73 \pm 0.02}\label{eqn:hiz_notcsfd}\\
A_{\mathrm{V,\,stacked}}(r) &= \left(2.3 \pm 0.5\right) \times 10^{-3} \left( \frac{r}{100\, h^{-1}\, \mathrm{kpc}} \right)^{-0.83 \pm 0.04}\label{eqn:stacked_notcsfd}
\end{align}
As above, Equations \ref{eqn:hidens_notcsfd}--\ref{eqn:stacked_notcsfd} describe the flatter parts of the profiles (50 \hinv kpc $ < r < 1$ \hinv Mpc). Finally, the SFD-based LRG profile is best fit by Equation \ref{eqn:lrg_notcsfd}:
\begin{equation}
A_{\mathrm{V,\,LRG}}(r) = \left(4.2 \pm 0.7\right) \times 10^{-3} \left( \frac{r}{100\, h^{-1}\, \mathrm{kpc}} \right)^{-0.95 \pm 0.04}\label{eqn:lrg_notcsfd}
\end{equation}
The normalization of Equation \ref{eqn:lrg_notcsfd} matches Equation \ref{eqn:lrg} within error bars, however, the exponents are in a moderately significant 3 $\sigma$ disagreement. Meanwhile, the normalization and exponent of Equation \ref{eqn:stacked_notcsfd} agrees with Equation \ref{eqn:stacked} within 1 $\sigma$.

\section{Additional Sources of Bias}\label{sec:other-systematics}

Leakage of the foreground sample into the background (or vice versa) would introduce a significant bias into our results and change the amplitude of the measured extinction. To minimize the probability of this sample leakage, we imposed relatively wide redshift separations: $\Delta z = $ 0.3--0.4 for the WISExSCOS–redMaGiC cross-correlation and $\Delta z = 0.15$ for the LRG sample. These separations exceed the WISExSuperCOSMOS normalized redshift uncertainty ($\sigma_z = 0.033$) and the redMaGiC uncertainty ($\sigma_z = 0.017/(1+z)$) by an order of magnitude.  


Even so, photometric redshift estimates are subject to catastrophic outliers and wide posterior distributions. To test for sample leakage in the face of these eventualities, we adopt a standard technique from weak gravitational lensing analyses and compute the boost factor profile $B(\theta) = 1 + w(\theta)$, where $w(\theta)$ is the angular clustering of galaxy positions. Lensing analyses use $B(\theta)$ to account for lens-source clustering when the two populations overlap in redshift \citep{2015ApJ...806....1M, 2022PhRvD.105h3528P,2004AJ....127.2544S}, making it a useful diagnostic for our purposes as well. 

\begin{figure} 
   \centering
   \includegraphics[width=0.75\linewidth]{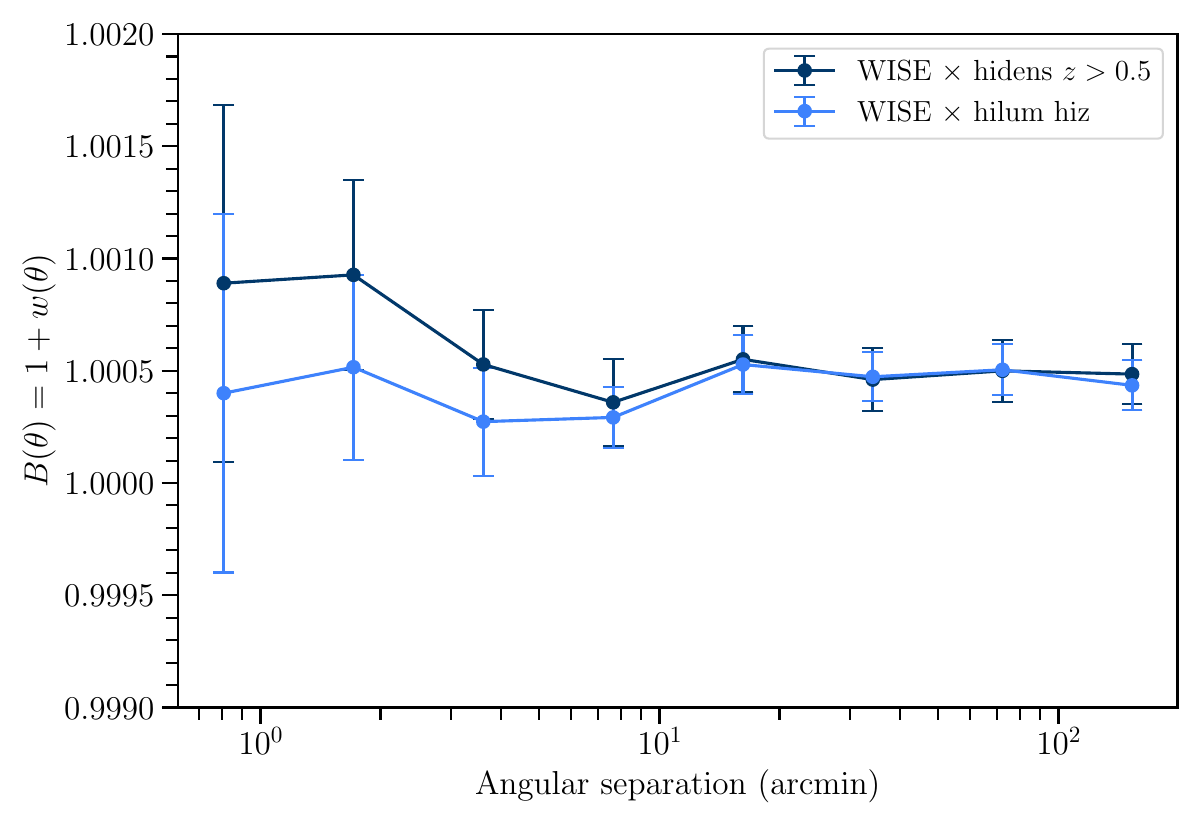}
   \caption{Boost factor profiles of redMaGiC background galaxies as a function of angular separation from foreground (WISExSCOS) galaxies.}
   \label{fig:density}
\end{figure}

We calculate $w(\theta)$ using TreeCorr's implementation of the Landy-Szalay estimator $\xi = (DD - DR - RD + RR)/RR$, which automatically accounts for survey geometry through the inclusion of random catalogs. Uncertainties are estimated from jackknife resampling across 50 sky patches. As shown in the left panel of Figure \ref{fig:density}, the resulting $B(\theta)$ for the cross correlation of foreground and background samples is consistent with unity at the $\leq 10^{-3}$ level over all angular scales considered. The corresponding $w(\theta) \simeq 10^{-4}$--$10^{-3}$ is one to two orders of magnitude lower than the clustering autocorrelation of redMaGiC galaxies themselves, e.g., \cite{2022PhRvD.105b3520A}. 
This result indicates that leakage-driven clustering between the foreground and background samples is negligible, and by extension, that foreground-background sample leakage is unlikely to be a significant problem in our analysis.


An additional, more subtle systematic concerns the role of extragalactic dust in defining background sample membership. If a background galaxy lies near a selection boundary in color or magnitude, dust-induced extinction and reddening can shift it across that boundary, changing which objects enter the extinction measurement. In this scenario, dust does not act solely as an additive color perturbation, but also as a sample selection operator: dusty sightlines may both redden galaxies and preferentially remove the reddest or faintest objects from the background catalog. If this ``sample selection reddening’’ is asymmetric---if more galaxies leave the background sample due to dust-induced reddening than enter it or vice versa---the mean color of the background sample will be shifted, changing the measured amplitude of the extinction signal by an unknown amount. 

This effect is inherently difficult to quantify, particularly for catalogs with complex, empirically calibrated selection functions like redMaGiC. The use of multiple background samples (i.e., redMaGiC high-density and high-lum high-$z$) with the same foreground sample (WISExSuperCOSMOS) provides a practical way to bound the magnitude of any such bias. Fitting each redMaGiC profile individually yields: 

\begin{equation}
A_{\mathrm{V,\,hidens}}(r) = \left(2.5\pm 0.1\right) \times 10^{-3} \left( \frac{r}{100\, h^{-1}\, \mathrm{kpc}} \right)^{-0.83 \pm 0.01}\label{eqn:hidens}
\end{equation}
%
\begin{equation}
A_{\mathrm{V,\,hiz}}(r) = \left(4.2 \pm 0.1\right) \times 10^{-3} \left( \frac{r}{100\, h^{-1}\, \mathrm{kpc}} \right)^{-0.95 \pm 0.05}\label{eqn:hiz}
\end{equation}

The normalization of the high-density background profile (Equation~\ref{eqn:hidens}) is approximately 40\% lower than that of the high-luminosity, high-redshift background (Equation~\ref{eqn:hiz}). Because the selection function of the high-luminosity high-redshift redMaGiC sample is similar---but not identical to---the high-density redMaGiC sample, 
we interpret the difference in normalization as a conservative estimate of the scale at which background sample selection can influence the inferred extinction amplitude. This difference is modest compared to the dynamic range of the measurement and does not alter the qualitative form of the extinction profile. More broadly, this exercise underscores the value of repeating extinction measurements for a given foreground sample with different, independently selected background samples, both as validation and as a means of bounding selection-related systematics.


\bibliography{main_rev2.bib}{}
\bibliographystyle{aasjournal}



\end{document}